# Anisotropic Crystallization Kinetics and Interfacial Dynamics of Phase-Change Material $Sb_2S_3$

Souvik Chakraborty, Wen-Qing Li, Yun Liu*

Institute of Advanced Intelligence and Computing (IAIC), Agency for Science, Technology and Research (A*STAR), 1 Fusionopolis Way, Level 16, Connexis North Tower, Singapore 138632, Republic of Singapore.

*Email: liu_yun@a-star.edu.sg

## ABSTRACT

Phase-change materials (PCMs) rely on rapid and reversible phase transitions between crystalline and amorphous states, which are critical for their performance in data storage, photonics, and thermal management applications. Antimony sulfide ($Sb_2S_3$), an Earth-abundant and sustainable chalcogenide, has emerged as a promising PCM due to its ultra-low optical losses, broadband transparency, and reversible phase-change capacity. In this work, we developed a machine learning force field based on the moment tensor potential approach, allowing us to understand the atomistic origin of the structural evolution and crystallization kinetics in $Sb_2S_3$ for the first time, by enabling large-scale molecular dynamics simulations (up to 7680 atoms for 40 ns). $Sb_2S_3$ has anisotropic growth rates with the (100) facet (*Pnma* space group) exhibiting the fastest growth due to the strong Sb-S covalent bonding along its quasi-1D ribbon-like structure of its crystalline phase. The maximum growth rate was observed at 875 K, with an optimal undercooling degree of ~100 K, significantly lower than other PCMs like $Ge_2Sb_2Te_5$ (GST) and GeTe. The activation energy for crystal growth was calculated be around 0.56 eV. Interfacial atomic dynamics revealed an Arrhenius-to-non-Arrhenius transition at ~800 K, with the (100) facet showing the highest diffusion activation energy of 1.56 eV. Unlike the GST and GeTe, the lower activation energy for crystal growth in $Sb_2S_3$ indicates that its heterogeneous crystallization is interface-controlled rather than diffusion-limited, with atomic attachment at the solid–liquid interface being energetically favoured over long-range atomic transport. Bonding and ring statistics analyses confirmed that directional covalent bonding plays a fundamental role in guiding structural ordering during crystal growth. These findings provide key insights into the structural, thermodynamic, and kinetic properties of $Sb_2S_3$, paving the way for optimizing its functionality including switching speed, reliability, and energy efficiency.

# 1. INTRODUCTION

Antimony sulfide ($Sb_2S_3$), an Earth-abundant chalcogenide, gained prominence as a versatile semiconducting material with wide applications such as phase-change memory[1], data storage[2], smart photonics[3–5], photovoltaic solar cells[6–9] etc. In particular, $Sb_2S_3$ exhibits robust phase-change behavior and thermal stability as it switches between the crystalline and amorphous phase.[10–14] It has broadband transparency from approximately 610 nm to the near-infrared region, unlike conventional phase change materials (PCMs) such as $Ge_2Sb_2Te_5$ (GST), which has substantial absorption losses across broad wavelength ranges. $Sb_2S_3$ also exhibits a refractive index contrast of $\Delta n = 0.60$ while maintaining extremely low extinction coefficients ($\Delta \mathrm{k} < 10^{-5}$) in the telecommunications C-band at 1550 nm.[15] The large $\Delta n$ together with low $\Delta k$ yields a high optical figure-of-merit (FOM = $\Delta$n/$\Delta$k), which is desirable for reconfigurable photonic applications.[16] The combination of ultra-low optical losses, reversible phase-change behaviour, strong absorption, and robust phase stability positions $Sb_2S_3$ as a key material for emerging optical and electronic technologies.

Insights into the structural, electronic, and phase behaviour of $Sb_2S_3$ have been investigated in the literature. For example, Kassem et al.[17] used a combination of high energy X-ray diffraction (XRD), Raman, differential scanning calorimetry (DSC) and *ab initio* MD to characterize atomic structure, dynamics and thermal and electrical properties of crystalline and glassy states of $Sb_2S_3$. Dong et al.[18] established $Sb_2S_3$ as a highly desirable candidate for active photonics applications. They showed that $Sb_2S_3$ exhibits a significant absorption edge red-shift (115 nm) upon crystallization; it has a high refractive index change ($\Delta$n ≈ 1 at 614 nm) and low extinction coefficient desirable for high-Q optical resonators and transmissive filters. $Sb_2S_3$ can be switched optically and electrically, with a crystallization time of approximately 78 ns and a re-amorphization time of 5 ns using laser pulses. $Sb_2S_3$-based photonic devices demonstrate large color shifts and spectral changes due to its absorption edge tuning. Hoff et al.[10] investigated the crystallization kinetics and glass dynamics of $Sb_2S_3$ with DSC techniques. Their study revealed that $Sb_2S_3$ crystallizes exclusively from the undercooled liquid (UCL) phase decoupling its crystallization process from the thermal history of the glassy phase, which in turn enhances repeatability and stability for non-volatile photonic devices.

However, experimental techniques struggle to probe crystallization kinetics across a broad temperature range and to resolve the underlying atomic-scale mechanisms. For all PCM applications, crystallization kinetics governs the phase transition dynamics, dictating the speed and fidelity of switching between amorphous and crystalline states. Experimental characterization of crystallization kinetics faces several intrinsic limitations such as resolving local nucleation events, transient intermediate phases, or capturing interface-driven growth that dominate in atomic scale. It is indeed difficult to capture the full crystallization trajectory under realistic device conditions. [19,20]

Machine learning force fields (MLFFs) have transformed materials research enabling rapid exploration of structural, dynamic, and thermodynamic properties across vastly larger phase domains than first-principles calculations permit.[21] Priyadarshini et al.[22] recently developed a MLFF for modelling $Sb_2S_3$ using the moment tensor potential (MTP) framework and an active learning approach. The trained MTP achieved high accuracy in predicting energy, forces, and stress tensors. The structural features were benchmarked with *ab initio* MD data and experimental results. Their MTP model effectively captured short and medium-range atomic order for amorphous $Sb_2S_3$. Fast and efficient computational methods are essential to access sufficiently long timescale while accurately capturing the rapid atomic rearrangements that occur during PCM crystallization. MLFFs can bridge this gap by delivering near first-principles accuracy with classical molecular dynamics (MD) speed.[23] Integrating MLFF with MD simulation makes exhaustive sampling of configurational space computationally feasible.[24–26] These advantages have been utilized to elucidate PCM properties in GST and GeTe, such as tracking crystallization in nanosecond timescale with growth rate estimation and interfacial dynamics across temperature ranges.[27,28]

Understanding the atomistic structure of the amorphous phase and the interfacial phenomena related to the phase transition is crucial for optimizing $Sb_2S_3$ performance in photonic computing and memory applications. To the best of our knowledge, a comprehensive study addressing the structural evolution during the crystal-to-liquid phase transition, the kinetics of facet-wise crystal growth rate, and the associated interfacial dynamics has not yet been reported in the literature. In this work, we present the development of a MLFF based on MTP and validate its accuracy through atomic-level structure analyses and existing experimental XRD results. The transformation of atomic configurations is examined to reveal the contrasting characteristics between the crystalline and liquid phases. Furthermore, we conduct a detailed investigation of facet-dependent crystal growth rates and evaluate the corresponding energetics and dynamics governing the crystal growth process.

**Table 1.** Comparison of $Sb_2S_3$ lattice parameters (Å). Unit cell structure and lattice vectors (*a, b, c*) are illustrated in Figure 1c.

| $Sb_2S_3$ | a | b | c |
|---|---|---|---|
| MTP (this study) | 3.878 | 11.089 | 11.428 |
| Experiment[29] | 3.836 | 11.229 | 11.311 |
| DFT (optB86b-vdW)[30] | 3.861 | 11.089 | 11.476 |

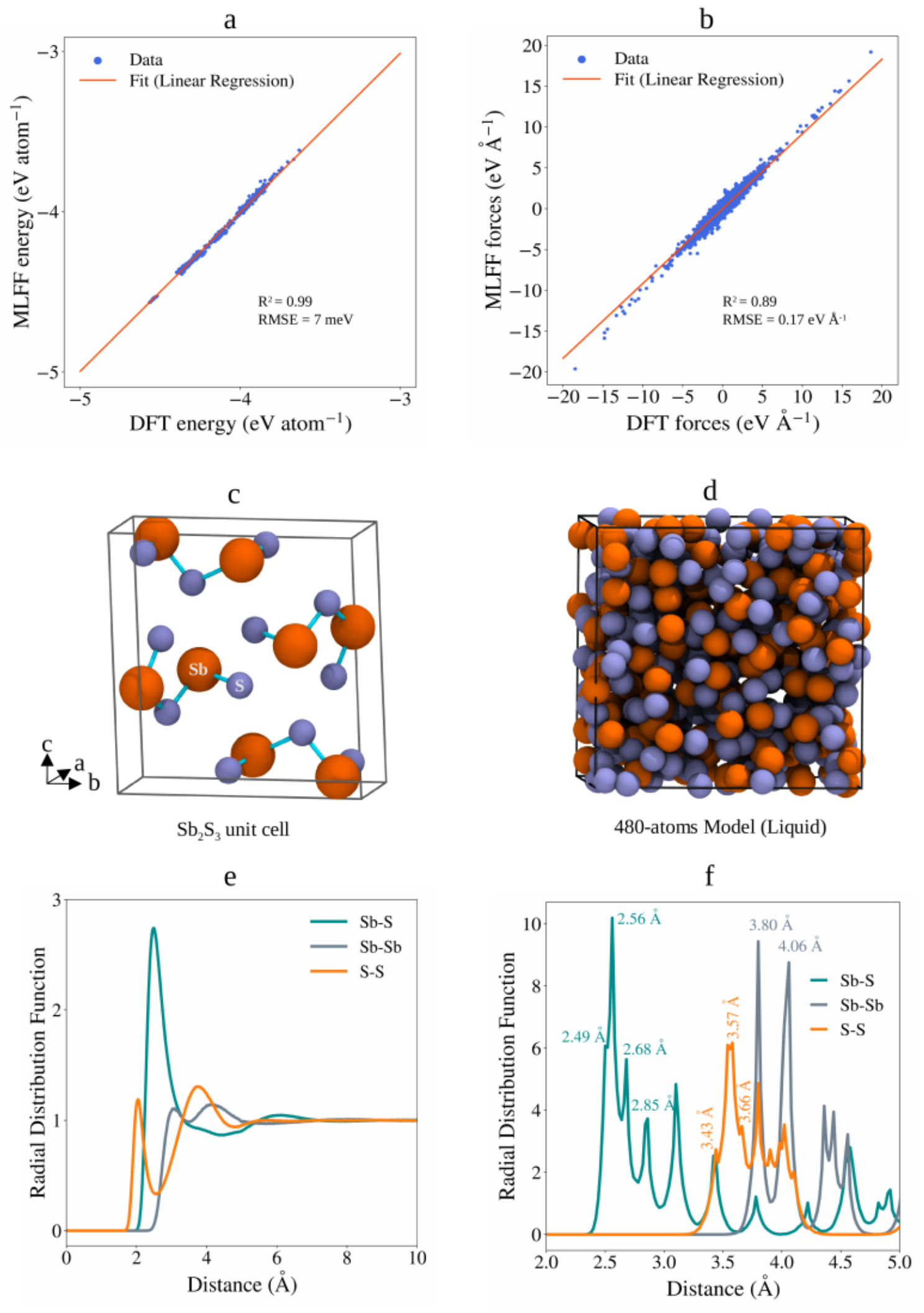

**Figure 1.** Correlation of the training set energies (a) and *x*-component atomic forces (b) obtained at DFT level and MTP prediction. c) The typical $Sb_2S_3$ unit cell structure in the *Pnma* space group. The orange and ice blue spheres represent Sb and S atoms, respectively. d) Snapshot of the equilibrated liquid phase structure of $Sb_2S_3$ at 3000 K. e) Radial distribution functions (RDFs) for all atom pairs of liquid $Sb_2S_3$ model at 3000 K. Interatomic distances were computed (see Table 2) from the first peak position in the respective RDF plot. f) RDFs of atom pairs for $Sb_2S_3$ crystal simulated for 0.1 ns at 300 K with MTP. The peak positions indicate the interatomic distances for respective atom pairs.

## 2. RESULTS AND DISCUSSION

### 2.1 Validation of MLFF

The root-mean-squared errors (RMSEs) of the training set for energy per atom and force components are 7 meV and 0.17 eVÅ$^{-1}$, respectively (Figure 1a and 1b). Our previous work demonstrated that the MTP obtained *via* the adopted scheme in this study can be stable and gives reliable predictions over long MD simulations on large and unseen amorphous structures.[31]

Accurately capturing low-level energy and force landscapes does not guarantee that emergent properties will be well reproduced. We therefore benchmark the MLFF-derived physical and structural metrics against experimental and *ab initio* results. We first optimized the $Sb_2S_3$ unit cell, which is illustrated in Figure 1c. The lattice parameters *a, b, c* were found to be 3.878, 11.089, and 11.428 Å, respectively (Table 1), which agree well with published first-principles DFT calculation and experimental values. The good agreements underscore the MTP model's accuracy and transferability.

Next, to validate the accuracy of the MTP model for amorphous $Sb_2S_3$, we performed MD simulation using a 480-atom model (Figure 1d) at 3000 K to obtain the liquid phase (details in the Models and Methods). The interatomic distances between the atoms (Sb-S, Sb-Sb, and S-S) were obtained from the 1$^{st}$ peak position of the radial distribution function (RDF) plots illustrated in Figure 1e. Table 2 shows the interatomic distances of all pairs of atoms obtained using our MTP model in good agreement with the reported experimental and *ab initio* values.[17] For crystalline $Sb_2S_3$, the RDFs and interatomic distances were obtained from a 0.1 ns MD run at 300 K using MTP (Figure 1f). The crystal phase interatomic distances match well with previously reported experimental results.[29,32] These agreements indicate that the developed MTP is sufficiently robust to be employed in subsequent large-scale molecular dynamics simulations of the crystallization process.

**Table 2.** Interatomic distance calculated from the position of 1$^{st}$ peak in RDF plots (Figure 1e) along with the values from high-energy X-ray diffraction (XRD) experiments and *ab initio* MD (AIMD) studies.

| **Interatomic Distance (Å)** | | | |
|---|---|---|---|
| | **Sb-S** | **Sb-Sb** | **S-S** |
| MTP (3000 K, this study) | 2.48 | 3.04 | 2.04 |
| XRD experiment[17] | 2.48 | 2.93 | - |
| AIMD[17] | 2.47 | 2.90 | 2.07 |

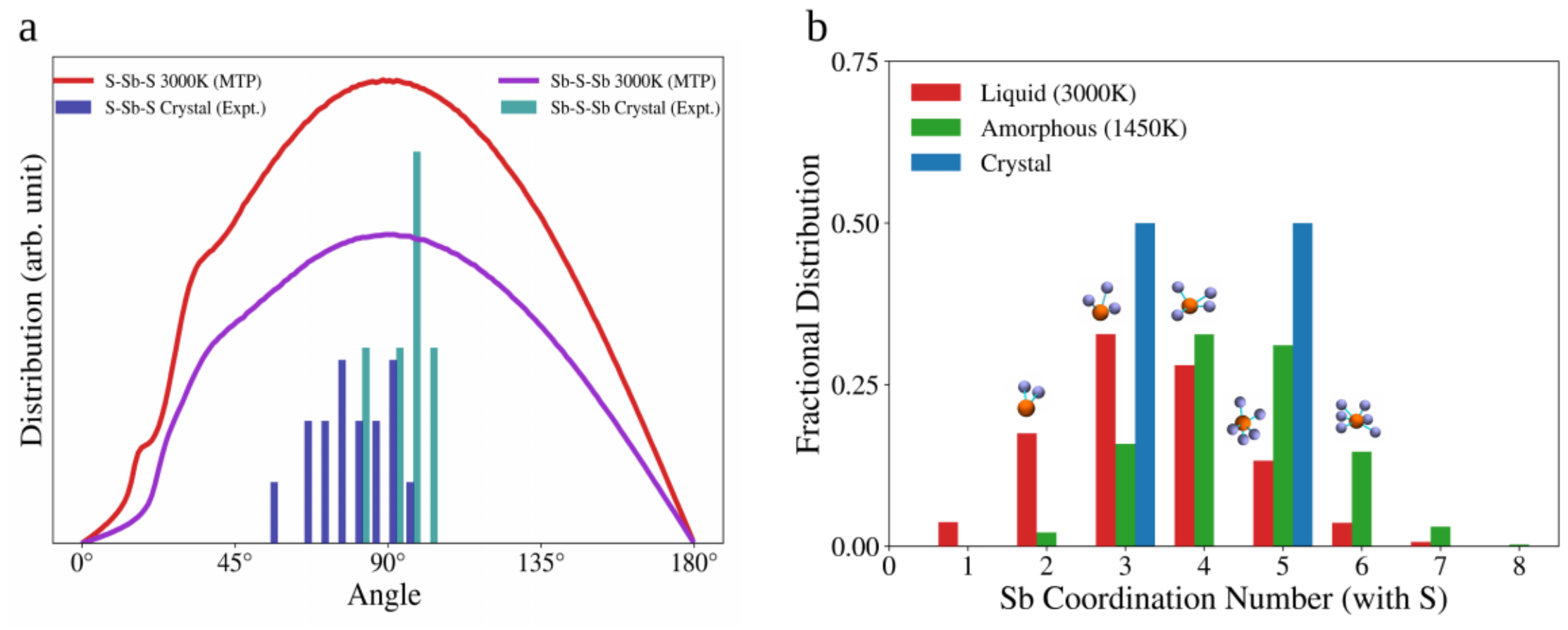

**Figure 2.** a) Angle distributions in crystal phase obtained by experimental [29] (X-ray diffraction) and liquid phase obtained by MTP in this work. b) Coordination number distributions (Sb with S) for crystal, amorphous and liquid phases. The characteristic structural motifs corresponding to the most prevalent coordination numbers in the liquid phase are illustrated.

## 2.2 Structural Characterization

To investigate how the geometry of cation ($Sb^{3+}$) and anion ($S^{2-}$) environments, orientational order, and configurational freedom differ between the crystalline and the liquid phase, we computed the angular distribution functions (ADFs) for the S–Sb–S and Sb–S–Sb bond angles, as shown in Figure 2a. The angles for crystalline phase are reproduced from previous experimental study[29]. For the ADF analysis, we used the Sb–S bond length corresponding to the position of the first minimum in the Sb–S RDF of the liquid i.e., 3.5 Å, as the cutoff for identifying angles formed by covalently bonded Sb and S atoms. The orthorhombic crystal structure of $Sb_2S_3$ contains two crystallographically distinct antimony sites: Sb(I) exists in a distorted trigonal pyramidal coordination with three sulphur neighbours, while Sb(II) adopts a slightly distorted square pyramidal coordination with five sulphur atoms.[32] S-Sb-S bond angles vary considerably due to the distorted coordination geometries. The Sb (I) site exhibits S-Sb-S angles ranging from approximately ~59° to ~98°. For Sb (II) site, the S-Sb-S angle ranges from ~69° to ~91°. Sb-S-Sb bond angles in crystal are reported to fall under the range ~84° to ~101°.[29] In our liquid phase, however, we observed that both S-Sb-S and Sb-S-Sb bond angle distributions have much broader distributions (0-180°) centred around 90° (Figure 2a). Similar angular arrangements were also observed from the amorphous phase (1450 K) angle distributions (Figure S1). Upon melting, the angular correlation is significantly reduced, which reflects the loss of directional order present in the crystalline stibnite structure. The broad distribution may be attributed to the dynamic and disordered arrangement of $Sb_2S_3$ units in liquid phase. Yet the observed dominant probability near 90° indicates that, even in the liquid, local motifs resembling distorted pyramidal or quasi-tetrahedral Sb–S coordination persist as energetically favourable arrangements, albeit without the rigid orientational constraints imposed by the crystalline lattice. The contrast in ADFs between crystalline and liquid $Sb_2S_3$ reflects structural adaptability that allows $Sb_2S_3$ to undergo rapid and reversible transformations between ordered and disordered states, a key requirement for high-performance PCMs. The persistence of local crystal-like environment in the liquid phase, as indicated by ADF broad peaks centred around 90°, further ensures that the phase transition involves local rearrangements rather than complete bond dissociation, enabling fast kinetics and excellent cyclability.

To further elucidate the local structural changes associated with the phase change, we calculated Sb(-S) coordination number in crystal, amorphous (1450 K) and liquid phase (3000 K). The computed fractional distributions are illustrated in Figure 2b. We selected Sb(-S) coordination as it primarily dictates the local geometry and structural distortions (e.g., loss of local order) upon phase transformation. For the crystal phase, a cutoff distance of 3.0 Å was selected, which was optimized by reproducing experimental coordination number at 3 and 5 for crystalline Sb(-S)[32], to identify the nearest-neighbour S atoms around Sb. However, we chose a 3.5 Å cutoff for the amorphous and liquid phases, corresponding to the first minimum in the radial distribution function (Figure 1e). We observed that Sb atoms exhibit different coordination geometries across the phases. The ordered structure observed for the crystalline $Sb_2S_3$ breaks down in the amorphous phase. Structural disorder at 1450 K broadens the coordination distribution: four- and fivefold environments become prevalent, with fractions of approximately 0.33 and 0.31, respectively, while some degree of three- and sixfold populations remain, approximately 0.16 and 0.15. At 3000 K, the liquid distribution shifts toward lower coordination, with dominant three- and fourfold sites contributing approximately 0.33 and 0.28. We also explored the Sb-Sb coordination distribution (Sb-Sb cutoff 3.5Å) in the amorphous and the liquid phase as depicted in Figure S1. Neither the amorphous nor the liquid phase exhibits significant structural ordering, apart from a small fraction of 2-fold and 3-fold coordinated motifs. This reduction and relative broadening in coordination number distribution reflects a transition from a rigid, directional framework to a more flexible, transiently bonded network. Preserving short-range motifs in liquid phase while retaining some degree of higher coordination value allows $Sb_2S_3$ switch between ordered and disordered states efficiently. For PCM performance, this change in local coordination is critical because it directly influences the optical contrast between phases, and the structural reversibility during cycling.

### 2.3 Simulating Crystallization Process with Different Facets at Crystal-Liquid Interface

The study of crystallization and growth kinetics of PCMs is fundamental to understanding and optimizing their performance in data storage and neuromorphic computing applications.[33,34] While traditional chalcogenide PCMs like $Ge_2Sb_2Te_5$ (GST) have been extensively characterized for their crystallization behaviour and switching dynamics[26,35–39], the orthorhombic crystal structure and one-dimensional chain-like atomic arrangement of $Sb_2S_3$ present distinct kinetic pathways that remain less explored. The MTP model allows us to carry out MD simulation with a large 7680-atom model and significantly longer timescale (40 ns) necessary to capture the crystallization and growth phenomena.

Figure 3a illustrates the crystal growth simulation using dual phase coexistence model at 700 K with (010) interfacial facet. The exact setup and details are described in the Models and Methods section. Figure 3b depicts the atomic arrangements at interfacial crystal plane for three facets considered in this study: (100), (010), and (001) planes. The growth of crystal phase was observed to start from the crystal-liquid interface normal to the interface plane. To estimate the crystal growth rate, we tracked the growth of crystalline slab by distinguishing the crystal-like atoms from the liquid ones (discussed in Supporting Information, Figure S2). The growth rate ($V_g$) was calculated as $V_g = dL(t)/dt$ where $L(t)$ is the effective (half) thickness of the crystalline substrate.[26] The effective (half) thickness is given by:

$$L(t) = \frac{N(t)}{2A\rho_n^{\text{crystal}}} \qquad (1),$$

where $N(t)$ is the number of crystal-like atoms, $\rho_{\text{n}}^{\text{crystal}}$ is the number density of crystal phase configuration obtained by energy optimization of crystal structure using MTP, $A$ is the cross-section area of simulation box plane normal to the growth direction, and the factor two is introduced at the denominator to account for the two growing surfaces. Figure S3 illustrates the growth of effective thickness as a function of simulation time at different temperatures. All facet-wise growth rates as a function of temperature can be found in Table S1. The plot of facet dependent growth rate ($V_g$) as a function of temperature is depicted in Figure 3c. Among the three facets, (100) showed the fastest growth velocity across the temperature range. At 600 K, the $V_g$ (0.61 $T_m^{MTP}$) was found to be quite slow (below 5 Å ns$^{-1}$). As the temperature increases, the undercooling decreases, and the growth rate increased almost linearly, and it reaches a maximum at 875 K for each facet (0.9 $T_m^{MTP}$). At 875 K, the $V_g$ for (100), (010), and (001) was 21.44, 12.93, and 7.67 Å ns$^{-1}$, respectively, indicating ~4 times faster with respect to the extreme undercooling limit of 600 K. $Sb_2S_3$ growth rate ($V_{g,\,max}$~2 m s$^{-1}$) was found to be slower than that of other PCMs like $Ge_2Sb_2Te_5$ ($V_{g,\,max}$~9 m s$^{-1}$) and GeTe ($V_{g,\,max}$~3.5 m s$^{-1}$).[26,40]

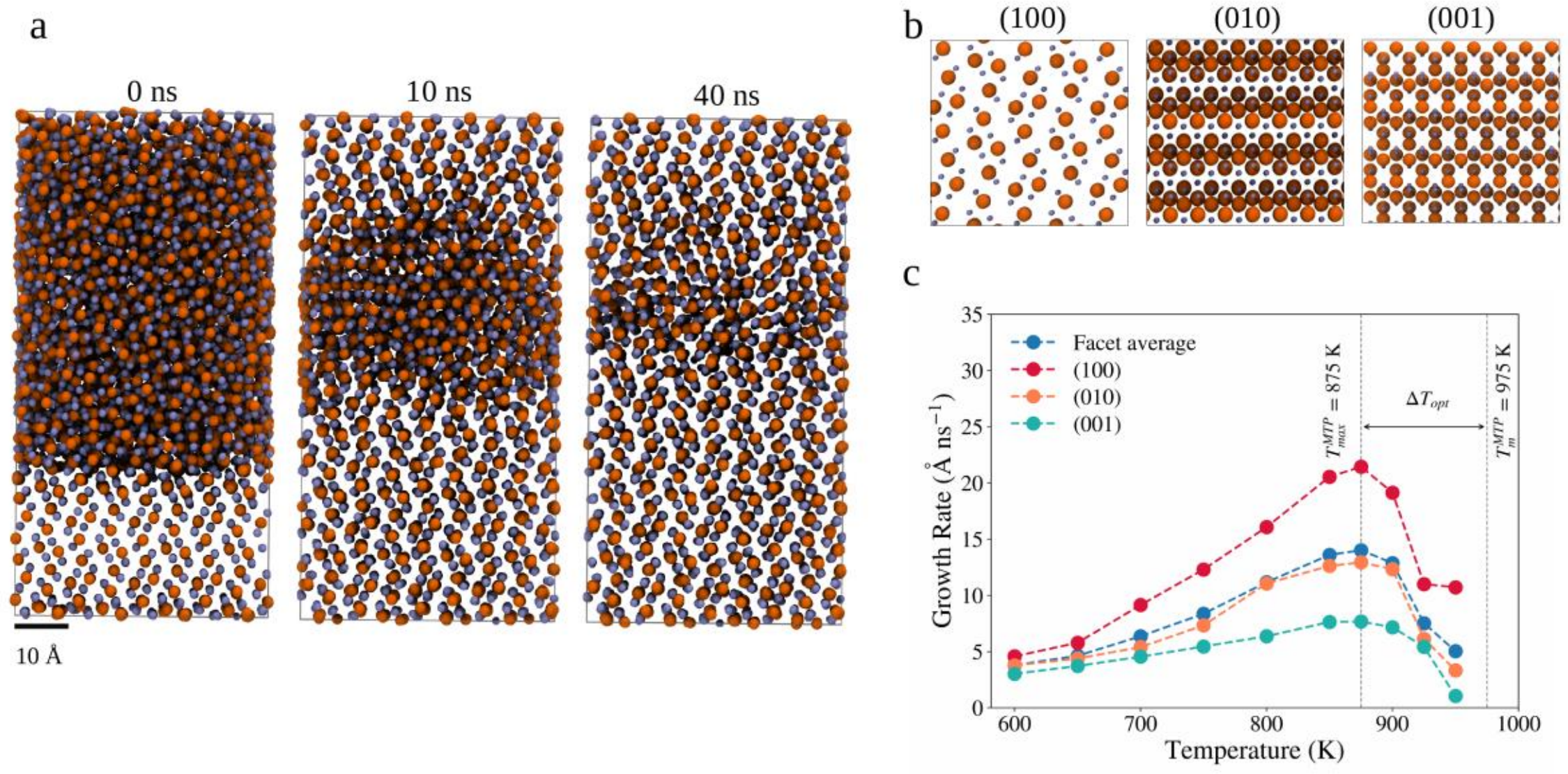

**Figure 3.** a) Crystal growth simulation of the dual phase coexistence (DPC) model with 7680 atoms. The evolution of crystal growth ([010] direction) is illustrated by the snapshots taken

from simulation trajectory (700 K) at 0, 10, and 40 ns respectively. The final equilibrated box dimension is 46.04 Å × 90.74 Å × 47.10 Å. For other facets, the growth direction remained parallel to the respective elongated box vector. b) Snaps of the three types of facets of $Sb_2S_3$ crystal at crystal-liquid interfaces considered in this study. c) The growth rate as a function of temperature for three types of interfacial crystal facets (100), (010), and (001) respectively. The circles represent the data, and the dashed lines are visual guides. The melting point temperature ($T_m^{MTP}$) obtained from MD simulation using MTP and the temperature of maximum crystal growth rate ($T_{max}^{MTP}$) are indicated by vertical dashed lines. The degree of optimal undercooling ($\Delta T_{opt} = T_m^{MTP} - T_{max}^{MTP}$) is pointed by the horizontal double-headed arrow.

We noticed that the difference in growth rates became pronounced after 700 K, and at 875 K, the maximum difference was attained (Figure 3c). At 875 K, growth for (100) facet was ~1.6 and ~2.8 times faster than that of (010) and (001), respectively. The fastest growth direction is the same direction that $Sb_2S_3$ crystal (a-axis, Figure 1c) forms needle-like motif with $(Sb_4S_6)_n$ units (Figure S4).[32] Experimental studies also showed that $Sb_2S_3$ grows anisotropically along the ribbon direction with aspect ratio increasing over time thereby creating an anisotropic bonding network.[41–43] Within a ribbon, there are Sb-S covalent bonds. The lateral interaction between the ribbons is much weaker and van der Waals in nature.[44] As strong bonds can continuously form along the ribbon direction it helps to reduce the surface energy to make the growth energetically favourable.[45,46] In addition, the stereochemically active lone pair ($5s^2$) of $Sb^{3+}$ contributes to an asymmetric coordination environment that promotes the formation of ribbonlike motifs.[44]

Close to melting point (875 < T < 975 K, small undercooling), the growth rate slowed as crystallization becomes unfavourable due to reduced supersaturation and increased thermal fluctuations at the interface. The melting point temperature ($T_m^{MTP}$) was computed by estimating enthalpy as a function of temperature (600–1200 K) as depicted in Figure S5. We found $T_m^{MTP} = 975\ K$, which implies that MTP overestimated the melting point temperature by ~17% compared to experimental value of 825 K[47–49] . Our simulation results qualitatively suggest that the maximum growth rate for $Sb_2S_3$ was achieved at a degree of optimal undercooling, $\Delta T_{opt}$ ( $T_m^{MTP} - T_{max}^{MTP}$, where $T_{max}^{MTP} = 875\ K$) of approximately 100 K. In general, low $\Delta T_{opt}$ is desirable for most PCM applications as it ensures better thermal responsiveness, effective energy usage, reduced thermal hysteresis, cycling stability etc.[50–53] For popular PCMs like $Ge_2Sb_2Te_5$ and GeTe, $\Delta T_{opt}$ can be found as ~270 K and ~210 K, respectively.[26,40] Our results show that $Sb_2S_3$ exhibits significantly lower degree of optimal undercooling at 100K.

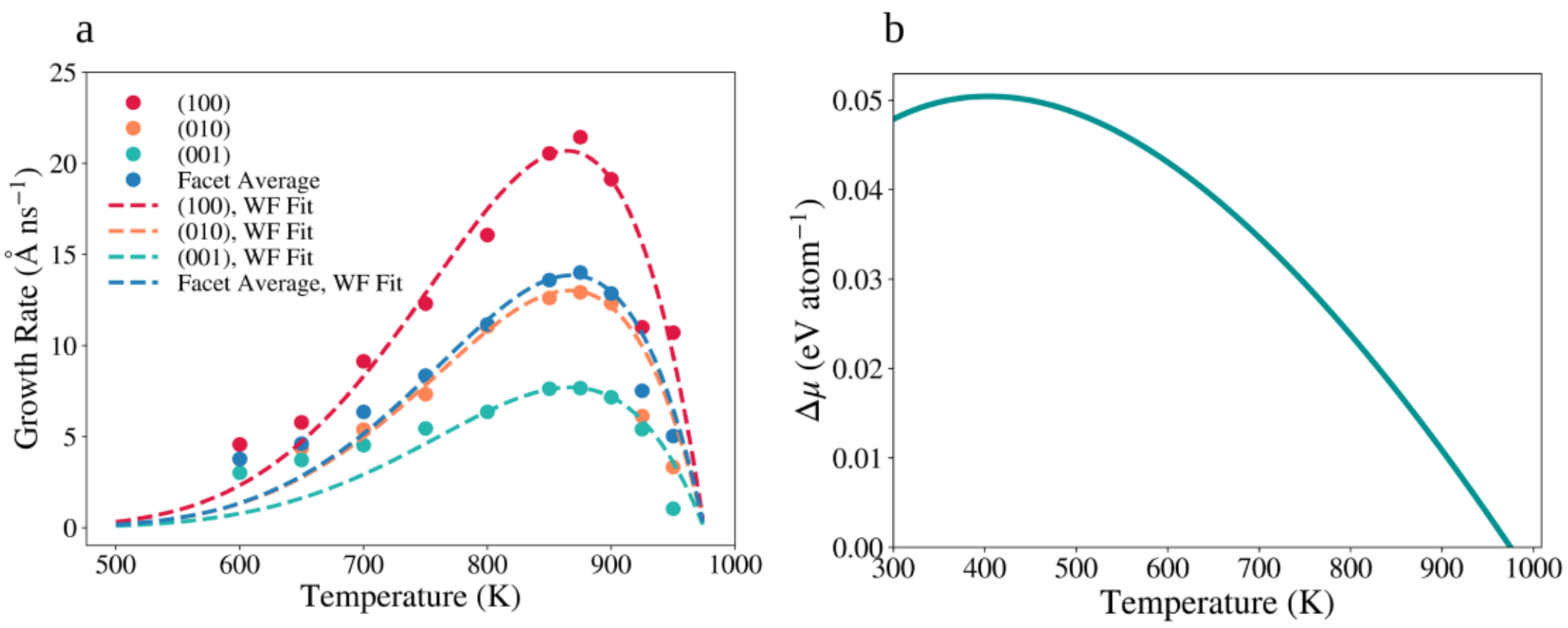

**Figure 4.** a) Wilson-Frenkel fitting (Equation 2) of the growth rate data for all three facets and the facet average. The fitting curves are represented by the dashed lines. From the fitting, the activation energy barrier for growth ($\Delta E_a^{WF}$) was computed. b) Free energy difference between crystal & liquid (Δμ) as a function of temperature from Thompson-Spaepen formula (Equation 3).

To describe the crystal growth rate as a function of temperature we fitted our results to the phenomenological kinetic model i.e., Wilson-Frenkel (WF) formula[54,55]

$$V_g = k_0 e^{\frac{-\Delta E_a^{WF}}{k_B T}} [1 - \exp(-\Delta\mu/k_B T)] \qquad (2),$$

where $k_0$ is the kinetic pre-factor, $\Delta E_a^{WF}$ is the activation energy for the crystal growth from the crystal-liquid interface, $k_B$ is the Boltzmann constant, Δμ is free energy difference between the crystalline and supercooled liquid phases. The WF formula is generally effective in describing a continuous growth of a rough crystal surface.[26,56–58] We illustrated the evolution of the rough growth front during crystallization from the interface in Figure S6. The free energy, Δμ is the thermodynamic driving force for crystallization. According to Thompson and Spaepen[59], Δμ is given by

$$\Delta\mu^{TS} = \frac{\Delta H_m(T_m - T)}{T_m} \frac{2T}{(T_m + T)} \qquad (3),$$

where the enthalpy of fusion, $\Delta H_\mathrm{m} = 0.147$ eV atom$^{-1}$, and $T_m$ is the melting point temperature from simulation ($T_m^{MTP} = 975\ K$) (Figure S5). Modifying the equation 2 by inserting the Thompson and Spaepen expression, we performed fitting on facet-wise data for crystal growth rate obtained at different temperature as shown in Figure 4a. The remaining parameters in equation 2, $k_0$ and $\Delta E_a^{WF}$ were determined by best fitting (goodness of fitting, $R^2$), and we obtained $\Delta E_a^{WF}$. $\Delta E_a^{WF}$ is the activation energy for crystal growth derived from simulation. Figure S7 illustrates the WF fitting plots for each facet and the facets-averaged data (illustrating data points considered for best fitting). Table S2 depicts the fitting parameters. The $\Delta E_a^{WF}$ values were found to be 0.554, 0.570, and 0.571 eV for (100), (010), and (001) respectively. The facet-averaged activation energy for crystal growth was obtained as 0.578 eV. The total crystallization activation energy $E_c$ is the sum of nucleation and growth activation energies. Experimentally obtained activation energy for crystallization ($E_c$) ranges from 1.31 to 2.72 eV[10]. In Table S3, we presented $E_c$ values for some important PCMs as obtained from the literature for comparison. Figure 4b depicts how Δμ varies with the degree of undercooling. Using equation 3 we see that Δμ increases with rising undercooling and eventually achieves an optimum value at around 400 K indicating significant effect of diminished atomic mobility due to arrested dynamics at extreme low temperature. Thus, when $\Delta\mu^{TS}$ expression was adopted in equation 2, WF model could capture the combined effect of kinetics and thermodynamics, providing us important properties associated with optimal crystal growth rate as shown in Figure 4a.

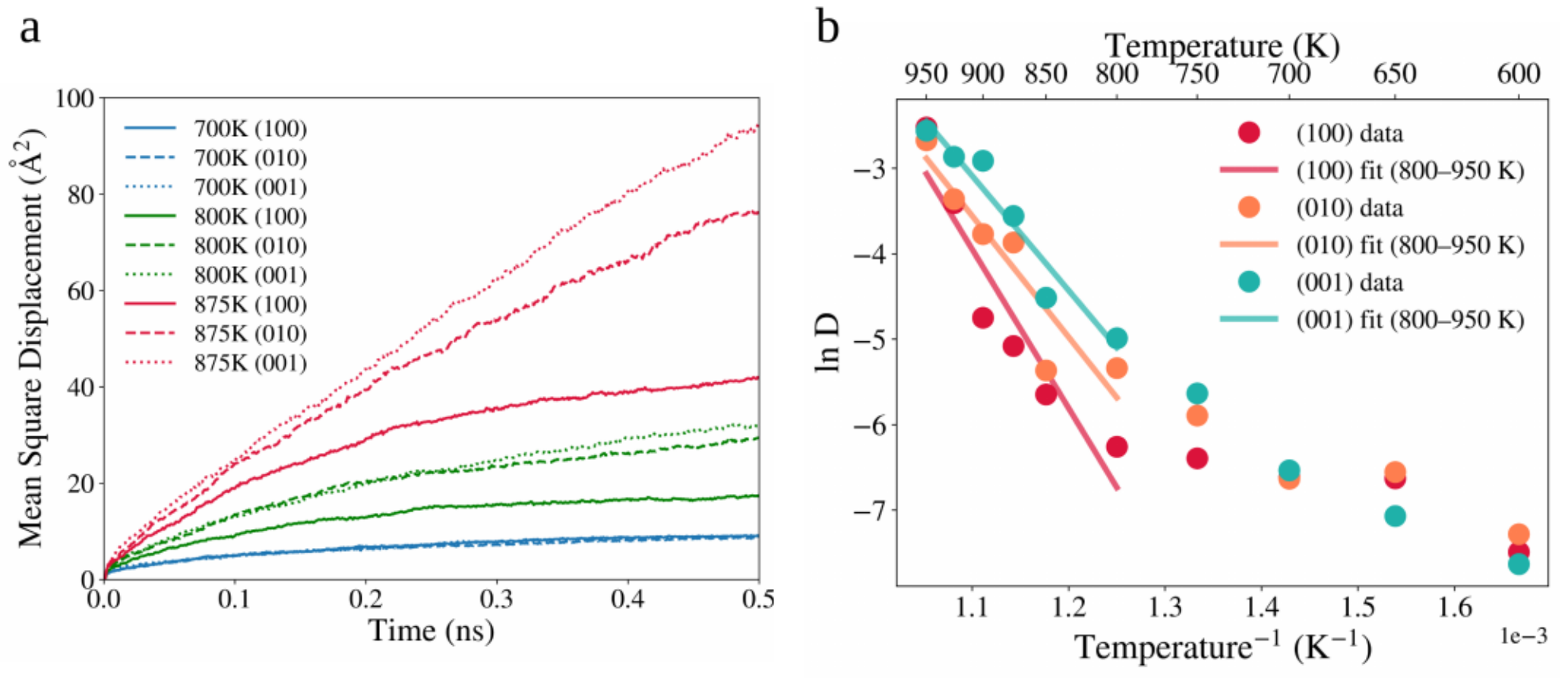

**Figure 5.** Effect of crystal growth on the dynamics of atoms within 1 nm (layer thickness) from the crystal slab at the onset of growth (0-0.5 ns) was evaluated by calculating mean-squared displacement (MSD) for each facet at all working temperatures. Self-diffusion constant (D) was calculated by linear fitting the MSD data (0.2-0.5 ns). a) MSD plots for 700, 800, 850K (chosen for illustration) are depicted for all three facets. b) Facet-wise Arrhenius fitting (equation 5) was performed on the diffusion constant as a function of temperature data by selecting the supercooled temperature regime (800-950K). Notice that Arrhenius nature deviates below 800 K. The Arrhenius activation energy for diffusion ($\Delta E_a$) was found to be 1.56, 1.24 and 1.16 eV for (100), (010), and (001) respectively.

### 2.4 The Effect of Crystal Growth on the Interfacial Atomic Dynamics

Now we turn to examine the effect of growth on the atomic mobility at the onset of crystal growth focusing on the interfacial liquid-like atoms. For that, we computed the mean square displacement (MSD) of atoms as a function of time for different temperatures (600-950 K). As we intend to elucidate the dynamics at the beginning timescale of the growth process, 0-0.5 ns of MD trajectory was considered for MSD calculations. We selected the liquid-phase atoms residing within 1 nm thick layer adjacent to crystalline slab in both directions. Figure 5a shows the mean square displacement (MSD) as a function of time for 700, 800 and 875 K (selected for illustration purpose) for three facets. From the relative slopes of the MSD curves as illustrated in Figure 5a, it is observed that increase in undercooling impacts the diffusive mobility of atoms as evident by gradual decrease in self-diffusion constant ($D$) as temperature decreases (Figure 5b). Self-diffusion constant is given by

$$D = \frac{1}{6} \lim_{t \to \infty} \frac{\partial \left( \frac{1}{N} \sum_{i=1}^{N} |\vec{r}_i(t) - \vec{r}_i(0)|^2 \right)}{\partial t} \qquad (4),$$

where the expression $\frac{1}{N} \sum_{i=1}^{N} |\vec{r}_i(t) - \vec{r}_i(0)|^2$ represents the MSD (averaged over the total number of atoms, $N$) estimated from the atomic position vector $\vec{r}_i(t)$. We computed $D$ from the slope of the linear part (0.2-0.5 ns) of MSD curves. All diffusion constant values are reported in Table S4. The Arrhenius equation for diffusion is given by

$$D = D_0 \, e^{\frac{-\Delta E_a}{k_B T}} \qquad (5),$$

where $D_0$ is the pre-exponential factor and $\Delta E_a$ is the activation energy for diffusion which represents the energy barrier for atoms to overcome to move from one site to another. In undercooled liquid, $\Delta E_a$ is the energy cost for atoms to break local cages or clusters formed by the neighbouring atoms which hinder the atomic movements impacting diffusivity. Figure 5b shows that the change in diffusivity with increasing undercooling is relatively faster in the range 950 to 800 K, and after that the rate of decrease slows until it reaches extreme undercooling limit at 600 K. 800 K (0.82 $T_m^{MTP}$) is found to be the point of deviation (from Arrhenius to non-Arrhenius transition with undercooling) for diffusion of atoms near the interface undergoing crystallization. The temperature range 800-950 K has been identified as the Arrhenius regime. At temperatures below 800 K, the system's fragility leads to noticeable deviations from Arrhenius behaviour. To calculate the activation energy $\Delta E_a$, we implemented linear fitting on the diffusion data at the Arrhenius regime for all facets. The activation energy for diffusion was found to be 1.56, 1.24 and 1.16 eV for (100), (010), and (001) respectively. If we consider the facet-wise impact on the dynamics of interfacial atoms undergoing crystallization, (100) facet shows the highest barrier followed by (010) and (001). We previously observed (100) facet offers the highest growth rate. It implies that locally atoms face difficulty to move through increasingly ordered and energetically conducive clusters formed due to simultaneous crystal growth event that arrests atomic thermal movement.

For heterogeneous crystallization of $Sb_2S_3$, the observation that the activation energy for crystal growth is less than the activation energy for diffusion i.e., $\Delta E_a^{WF} < \Delta E_a$ implies that $Sb_2S_3$ crystallization is not purely diffusion-controlled but rather interface-controlled, where local bond rearrangement at the solid–liquid interface dominates the kinetics. In such scenario, the structural motifs in the liquid phase undergo faster configurational adjustments to complete ordering at the interface than they diffuse through the melt. This makes atomic attachment energetically favourable than long-range transport in the bulk. This finding is consistent with the experimental observation reported in literature.[60] In contrast, the studies on GST and GeTe demonstrated that crystal growth in these phase-change materials is predominantly diffusion-limited.[61–63]

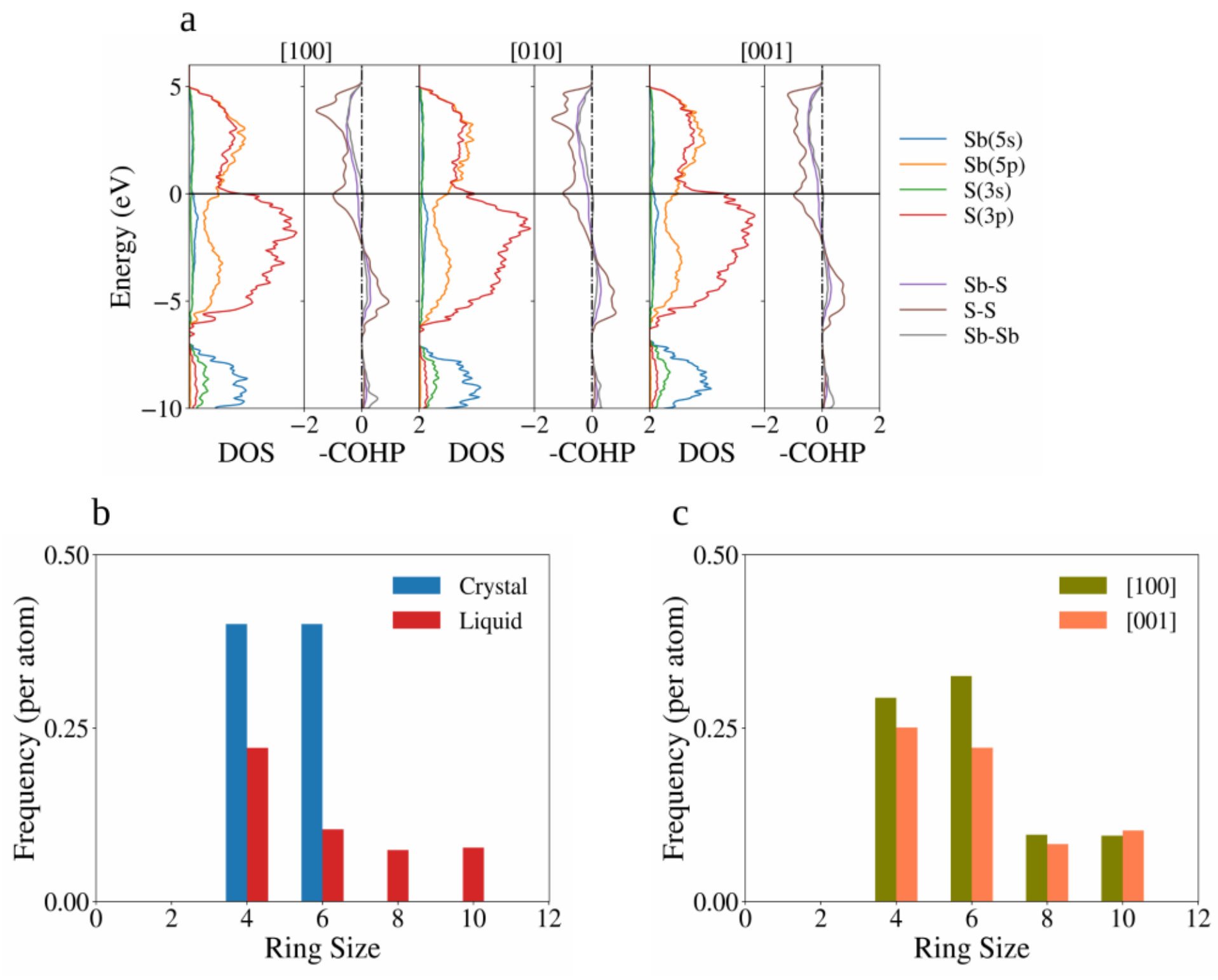

**Figure 6.** a) PDOS and -COHP analysis of the crystallization interface along the [100], [010] and [001] growth directions. b) Ring size distribution for primitive rings formed by Sb-S bonds: crystalline and liquid $Sb_2S_3$ system (3000 K) c) Comparison of ring size distribution for 1 ns configuration of interface systems with [100] and [001] growth directions at 875 K. The interfacial configurations were obtained from respective growth simulation trajectories.

To further understand the interfacial dynamics, we have performed crystal orbital Hamiltonian population (COHP) analysis[64] as implemented in the LOBSTER code[65,66]. As a reference, we first compared the bonding behaviour of crystalline $Sb_2S_3$ with $Sb_2Te_3$, another common PCM. The projected density of states (PDOS) and corresponding COHP analysis show that $Sb_2S_3$ and $Sb_2Te_3$ exhibit very similar bonding characteristics (Figure S8). At around -9.0 eV with reference to the valence band maximum (VBM), there is strong bonding interaction between Sb(5s) and S(3s)/Te(5s) orbitals. Between -5.0 eV to -2.0 eV, there is also bonding interaction between the Sb(5p) and S(3p)/Te(5p) orbitals. The states near the VBM are mainly composed of antibonding interactions between Sb(5s) and S(3s)/Te(5s) orbitals. The conduction band near the bandgap are mainly composed of antibonding interaction between Sb(5p) and S(3p)/Te(5p) orbitals. We note that other bonding interactions such as the homopolar S-S and Sb-Sb bonds are negligible from the COHP analysis.

We extend the COHP to the three different crystallization directions. To make the DFT calculation computationally tractable, we constructed smaller dual-phase coexistence models with a total of 480 atoms using partial melting at 3000 K. After equilibration with MD runs (1 ns NPT) with MTP, we performed DFT single point calculations on a snapshot of these structures without further relaxation. As seen in Figure 6a, the PDOS are very similar between the three growth directions. COHP analysis reveals strong homopolar S-S interactions in the liquid phase of $Sb_2S_3$, which are important at controlling the rate of charge transfer and the rate of disorder[67]. They exhibit strong bonding characteristics at around -5.0 eV below the Fermi level, and strong antibonding characteristic near the Fermi level and in the conduction bands. The S-S bond strengths are similar to those homopolar Ge-Ge and Ge-Sb bonds in the GST family of PCM[68]. In contrast, the Te-Te homopolar bonds are very weak in GST. In addition, the S-S COHP analysis exhibits the sharp peak at the -5.0 eV range (bonding interactions), and the 4.0 eV energy range (strong antibonding interactions), along the [100] crystal growth direction. This gives further evidence to our finding that along the [100] crystallization direction, strong covalent bonds result in the fastest crystal growth rate, compared to the [010] and [001] directions.

To understand how atomic attachment facilitates the crystal growth from interface we calculated ring size distribution for crystal, liquid and two interface systems having [100] and [001] growth directions with configurations at 1 ns of growth trajectory at 875 K. For that, we considered primitive rings formed by Sb-S covalent bonds[69,70]. For the ring statistics analysis, the Sb-S cutoff for crystal and interfacial configurations was set as 3 Å to consider intra-ribbon Sb-S bonds, and for liquid, it was 3.5 Å (~1$^{st}$ minimum from RDF). The figures for ring size distribution are illustrated in Figure 6b. $Sb_2S_3$ crystal has 4 and 6-membered rings whereas in liquid phase the bigger entities such as 8 and 10-membered rings appeared along with 4 and 6-membered rings. We observed that although the population of 4 and 6-membered rings has diminished in liquid phase (Figure 6b), yet they are the dominating motifs analogous to crystal phase. Similar trend was reported for GST[71]. Figure 6c demonstrates the comparative ring statistics between the fastest crystallization facet (100) and the slowest one (001) considering the interfacial configurations at 1 ns. We observed that in case of the fastest crystallization ((100) facet) 4 and 6-membered rings have higher population than the facet with slowest crystal growth ((001)). The higher 4 and 6-membered ring population in case of (100) facet is a result of favourable Sb-S covalent bond formation along the growth direction.

During the crystal growth, the complex interplay and competition between atomic diffusion and the driving forces for crystal formation such as activation energy for growth determine the optimal growth pattern. Different crystal facets at crystal-liquid interface have unique atomic arrangements and bonding environment. It leads to different kinetic, thermodynamic and diffusive variations. For efficient PCM design, the tuning of optimal facet may potentially play a crucial role depending on the desired function.

## 3. CONCLUSIONS

In summary, this work presents a comprehensive investigation of the mechanism and kinetics associated with heterogeneous crystallization of $Sb_2S_3$, a critical phase-change material (PCM) by developing a machine learning force field (MLFF) and consequently integrating with molecular dynamics simulation. Structural analyses through radial distribution functions, coordination numbers, and angular distributions distinctly revealed the contrast between the crystalline and liquid phases, confirming the MLFF's efficacy in capturing microscopic changes of short-range order and long-range disorder due to phase transformation. The structural descriptors helped to understand how local atomic rearrangements underpin transformation and influence the interfacial behaviour during the crystallization process.

Crystal growth simulations for the (100), (010), and (001) facets demonstrated clear anisotropy in growth kinetics. Quantitative kinetic and dynamic evaluation of facet-dependent growth rates and their corresponding activation energies indicated that interfacial atomic attachment and diffusion dynamics are strongly orientation-dependent, governed by variations in bonding configuration and diffusion barriers. At the atomistic scale, diffusion behaviour at the crystal–liquid interface elucidated the coupling between local mobility and growth front propagation, offering a molecular level mechanistic view of the crystallization process. COHP and ring statistics analyses further confirmed the underlying role of covalent bonding in driving directional crystal growth. By elucidating the energetics of crystal growth and diffusion, we infer that $Sb_2S_3$ crystallization is dominated by the local ordering at solid-liquid interface.

These insights help to establish a direct connection between microscopic atomic dynamics and macroscopic growth behaviour, offering predictive capability for PCM crystallization kinetics. For large-scale simulations of phase transitions, the developed MLFF-MD workflow provides an efficient framework with near first-principles accuracy. The understanding acquired from the outcomes of this study can inform and further advance the design of PCM materials with optimized switching speed, controlled crystallization pathways, and improved reliability for data storage and thermal management technologies.

## METHODS

### *MLFF Development*

To develop a reliable machine learning (ML)-based potential, we employed a bootstrapping technique based on moment tensor potential (MTP)[72] and density functional theory (DFT) calculations[73]. This approach assembles the training set by sampling configurations directly from the molecular dynamics (MD) simulations, starting with one crystalline (Pnma) and 20 initial non-crystalline structures. The bootstrapping method is an active-learning procedure that terminates simulations based on extrapolation criteria, retrains the potential, and restarts the simulation.[31]

Given 21 initial structures, this iterative loop continuously incorporates newly generated configurations, which are labelled by DFT, until a converged MTP model is achieved for a 50 ps MD simulation of each structure. The final training dataset comprises 2,094 structures, using an MTP of "level 12" with a cutoff of 5 Å, a minimum interatomic distance of 1.5 Å, and a radial basis size of 12. Here, the 20 initial non-crystalline structures were constructed by randomly blending and packing 24 Sb and 36 S atoms into a supercell using the initialization scheme proposed by L. B. Vilhelmsen and B. Hammer[74]. Each system was initially set up at low density within a large supercell, then compressed to an appropriate volume over 5 ps at 10 K under a high pressure of 50,000 bars in an NPT ensemble. Subsequently, an optimization step was performed to relax atomic positions and internal stress. These calculations utilized a universal machine learning force field implemented in MACE[75]. During the extrapolation step, the configurational space was further sampled through NVT and NPT simulations at varying temperatures (200–3500 K) and pressures (1–100 kbars).

DFT calculations were conducted using the Vienna Ab initio Simulation Package (VASP 5.4.4)[76,77]. The projected augmented wave method and the Perdew-Burke-Ernzerhof (PBE)[78] generalized gradient approximation were employed for the exchange-correlation functional with the vdW dispersion correction (Grimme's D3)[79]. An energy cutoff of 520 eV was used and an automatically generated k-mesh with the number of k-points along each reciprocal lattice vector was determined by setting a minimum spacing of 0.4 $Å^{-1}$ between k-points.

***Molecular Dynamics (MD) Simulation***

For the study of the amorphous/liquid phase, we constructed a 480-atom model of $Sb_2S_3$ was constructed by generating a cubic supercell (6×2×2) from its crystallographic unit cell. To achieve the well-equilibrated liquid phase $Sb_2S_3$ model, we minimized the total energy of the initial structure and then equilibrated the configuration at 3000 K for 12 ns in the NPT ensemble using MTP. To prepare the amorphous model, the liquid system was quenched to 1450 K at a cooling rate of 10 K $ps^{-1}$, followed by 5 ns of equilibration in the NPT ensemble. For facet-dependent crystal growth studies, we prepared 7680 atoms systems representing Dual Phase co-existence (DPC) model. First, rectangular supercells were constructed by replicating the unit cell along the principal lattice directions. The corresponding dimensions of the resulting supercells elongated along the a, b, and c axes were 24×4×4, 12×8×4 and 12×4×8, respectively. After performing the energy minimization of each supercell structure, the partial melting was employed to create the crystal-liquid interface (DPC model). The atoms in the part of the box (>25 Å of the elongated box vector) were heated at 3000 K for 2 ns (NVT) and the rest was preserved in crystalline form by implementing position restraint on atoms. By this method, we achieved three different DPC model configurations with three different crystal planes (100), (010) and (001) at the crystal-liquid interface. Then, each system was quenched to $T \in \{600 - 1200\ K\}$. After quenching, the position restraint was removed and finally, the system was equilibrated for 40 ns with NPT at each temperature (T). At 1000 K, the crystal-liquid interface disappeared, indicating that the substance has fully melted. Below 1000 K, we observed the crystal growth phenomena starting from the crystal-liquid interface. Identification of solid atoms is crucial to estimate the crystal growth rate. For that, bond orientational order parameters were widely considered e.g., Steinhardt order parameter.[80]

All MD simulations were performed with the LAMMPS-MLIP-3 interface package[81,82]. At first, the system was energy minimized with conjugate gradient method (implemented in LAMMPS[82]) to eliminate unfavourable atomic overlaps in the model. The MD simulation was executed with the integration time step of $5\times10^{-4}$ ps. For quenching, NVT ensemble was used with Nosé -Hoover thermostat[83,84] with a damping parameter 0.1 ps. For equilibration steps, NPT ensemble was implemented with Nosé–Hoover barostat and thermostat methods. The equilibrium pressure was set as 1 bar with the damping parameter of 1 ps.

## Author Contributions

Y.L. conceived and supervised the project. S. C. performed all the MD simulations and most the data analysis. W. Q. performed the DFT calculations and trained the MTP model. S.C. wrote the original draft of the manuscript. All authors discussed the results and commented on the manuscript at all stages.

## Conflicts of Interest

The authors declare no conflict of interest.

## Data availability

The MD simulation input files, including initial coordinate files, force-field file, and simulation parameters, are available in Zenodo at https://zenodo.org/records/20798209.

## Acknowledgements

Y.L. acknowledges funding support from A*STAR under its Young Achiever Award. This work is supported by A*STAR Computational Resource Centre (ACRC) and National Supercomputing Centre (NSCC), Singapore, through use of their high-performance computing facilities.

## Table of Contents

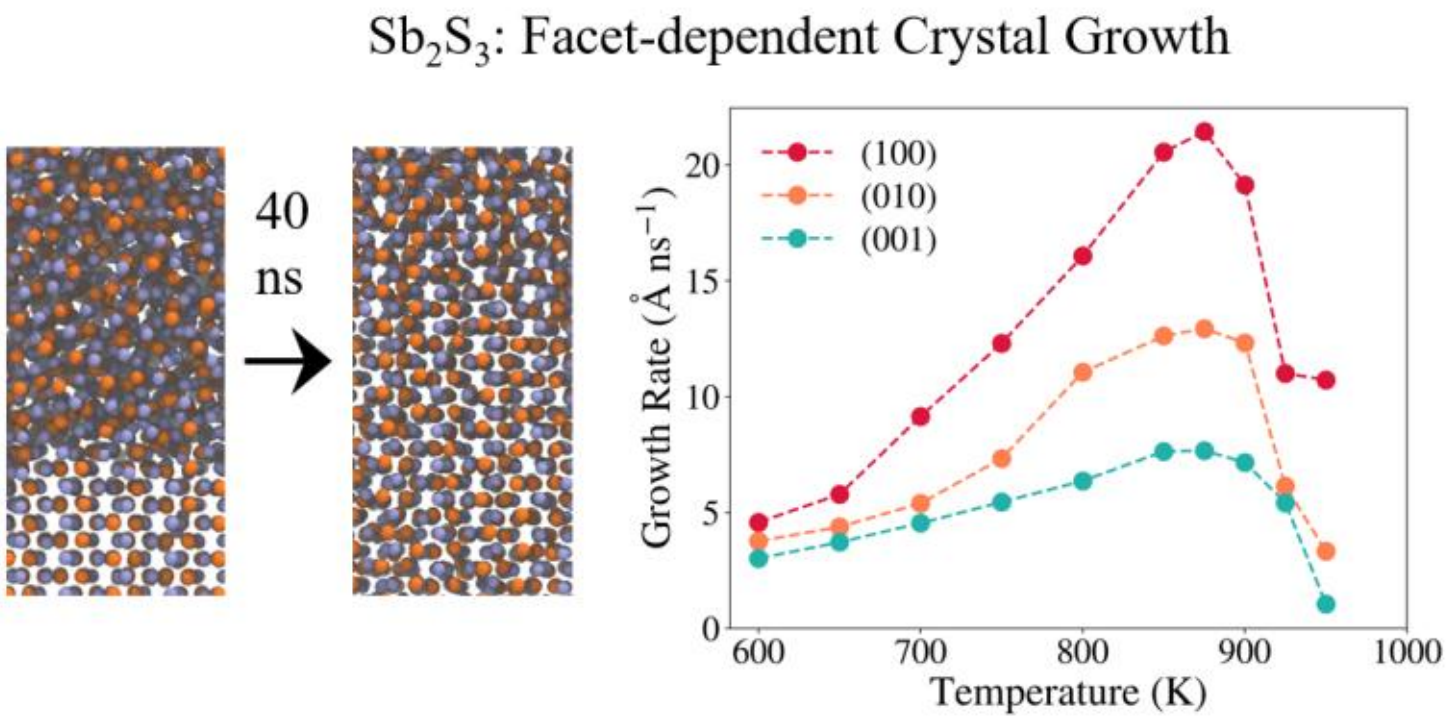

Antimony sulfide, a crucial phase change material, undergoes heterogeneous crystallization with different crystal facets. The study finds its crystal growth is highly directional primarily controlled by atomic interfaces, rather than atomic mobility.

# Supplemental Information for

# Anisotropic Crystallization Kinetics and Interfacial Dynamics of Phase-Change Material $Sb_2S_3$

Souvik Chakraborty, Wen-Qing Li, Yun Liu*

Institute of Advanced Intelligence and Computing (IAIC), Agency for Science, Technology and Research (A*STAR), 1 Fusionopolis Way, Level 16, Connexis North Tower, Singapore 138632, Republic of Singapore.

*Email: liu_yun@a-star.edu.sg

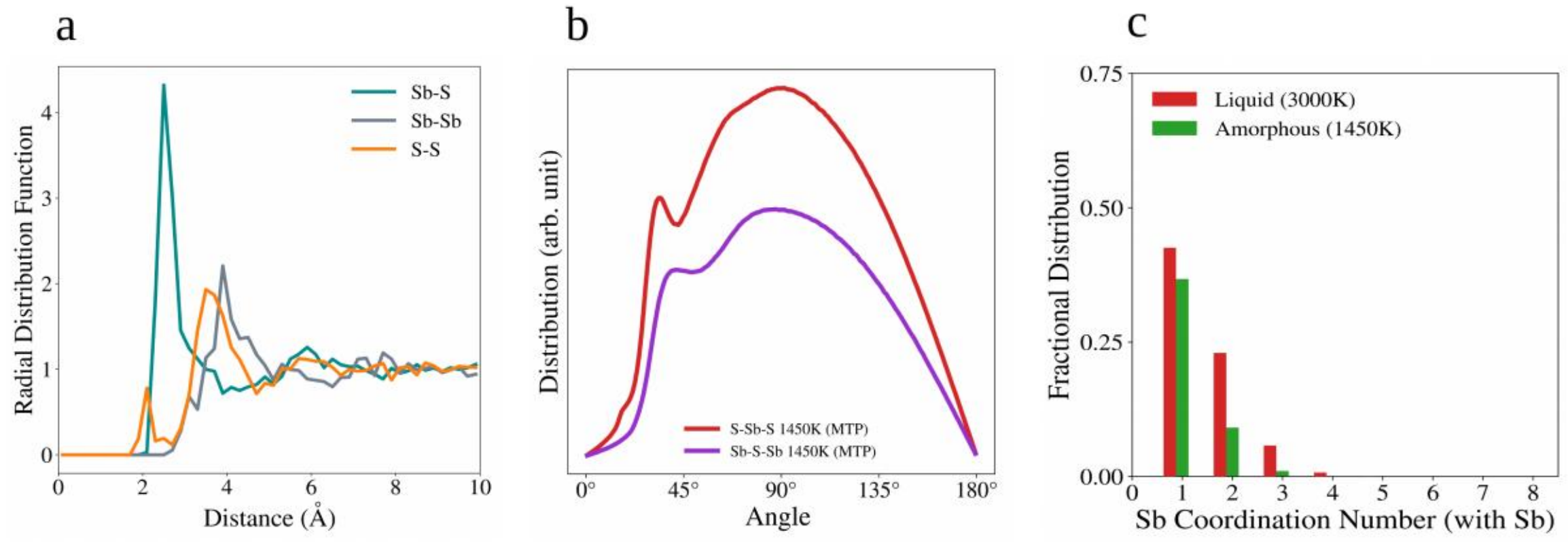

**Figure S1.** a) Radial distribution function plots for amorphous $Sb_2S_3$ at 1450 K. b) Angle distribution function (1450 K). c) Sb-Sb coordination number distributions in the liquid and amorphous phase.

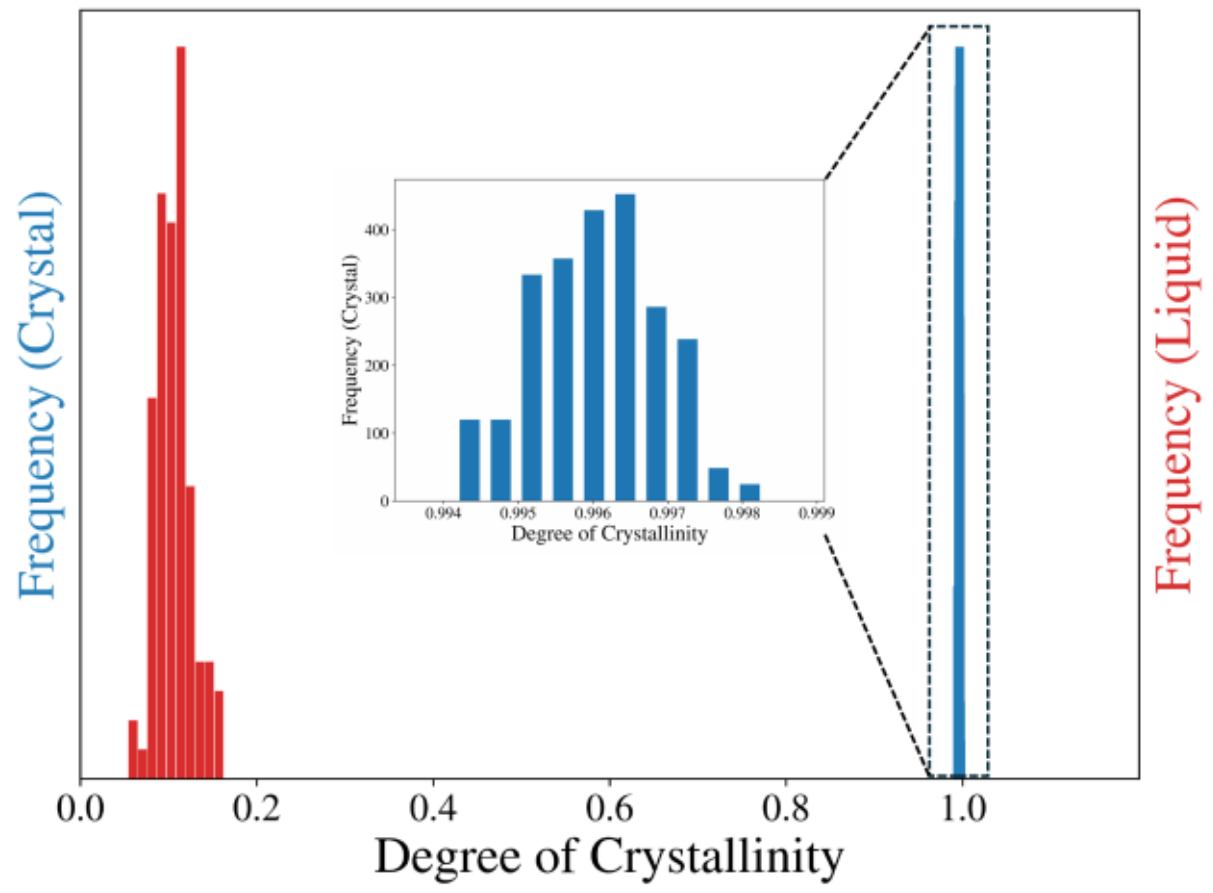

**Figure S2.** The degree of crystallinity was defined as the ratio of the number of solid atoms and the total number of atoms. To identify the solid atoms, we adopted $S_{ij}$ parameter[1,2] and implemented for this study using pyscal library[3]. $S_{ij}$ was calculated from the Steinhardt order parameter ($q_4$). The criteria for classifying solid atoms are:

$$S_{ij} = \sum_{m=-4}^{4} q_{4m,i}\, q^{*}_{4m,j} \geq \text{threshold}$$

$$\langle S_{ij} \rangle > \text{avgthreshold}$$

The order parameter $q_4$ is given by:

$$q_{4m,i} = \frac{1}{N_i} \sum_{j=1}^{N_i} Y_{4m}(\mathbf{r}_{ij})$$

Where $Y_{4m}$ is the 4$^{th}$ order spherical harmonic with order $m$, $\mathbf{r}_{ij}$ is the connecting vector between atom $i$ and $j$. $N_i$ is the number of neighbours within a given cutoff (4.5 Å in this study). To optimize the threshold and avgthreshold values for our system, we calculated and compared the distribution of the degree of crystallinity for pure solid and liquid systems. For that we considered the time averaged (last 100 frames) values of pure crystal model at 300 K and pure liquid model at 3000 K computed from their respective MD trajectory. We found that with threshold = avgthreshold = 0.22, the degree of crystallinity histogram curves (pure crystal vs. pure liquid) was found well-separated, as depicted in the figure. The crystallinity for pure crystal was obtained close to 1 meaning almost all atoms were identified as solid, whereas for the pure liquid it was found to be less than 0.2. It indicates the distinction of solid and liquid atoms was robust with the adopted thresholds.

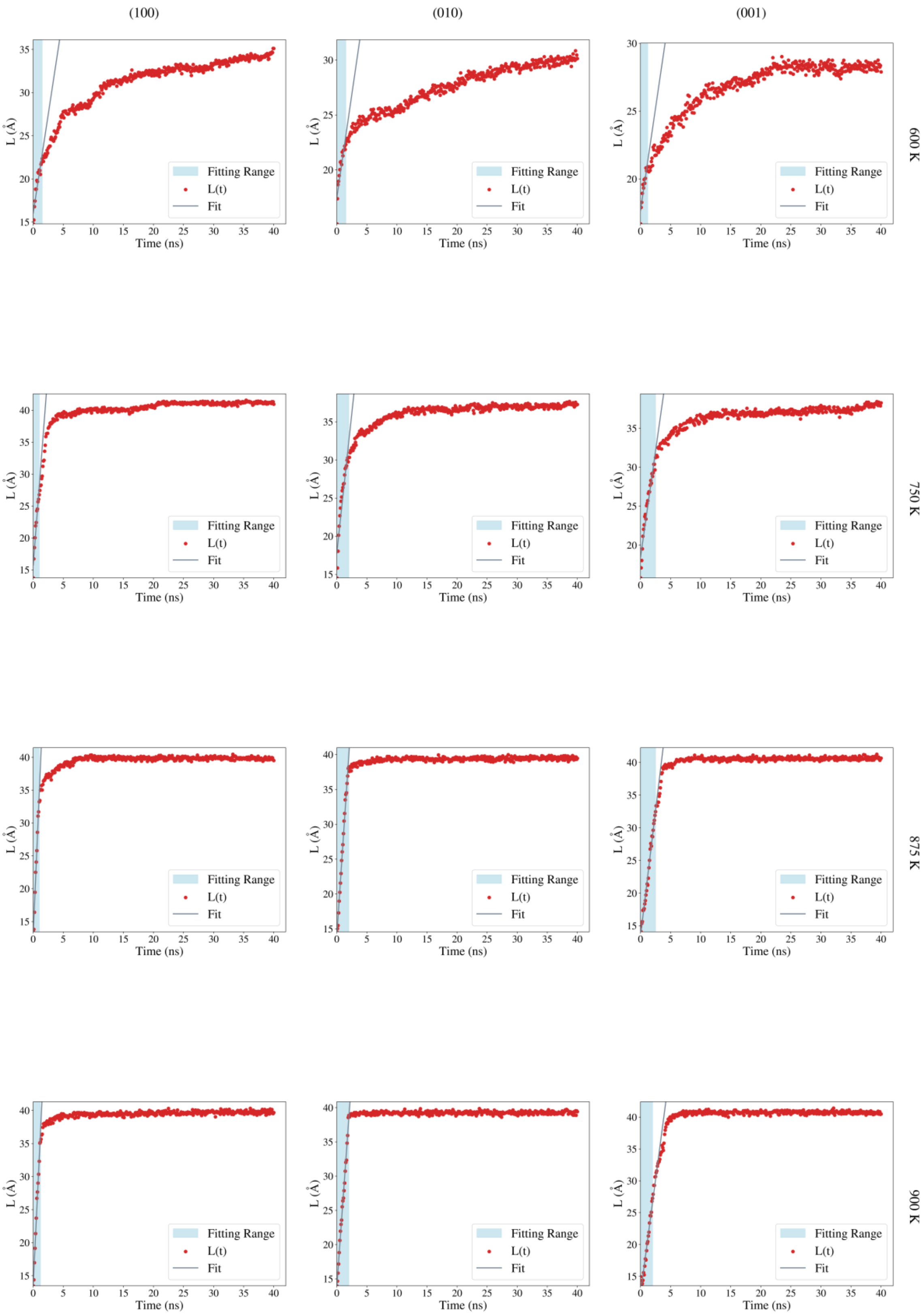

**Figure S3.** Effective-half-thickness (*L*) plots with linear fitting for three types of facets shown for T(K) = {600, 750, 875, 900}. The range considered for linear fitting is highlighted in light blue. The growth rate was estimated from the slope of the fitted line.[4] The detailed method is discussed in the main article.

**Table S1:** Growth rates (Å $ns^{-1}$) obtained for different facets at various temperatures (600-950 K).

| Facet | 600 | 650 | 700 | 750 | 800 | 850 | 875 | 900 | 925 | 950 |
|---|---|---|---|---|---|---|---|---|---|---|
| **(100)** | 4.580 | 5.783 | 9.146 | 12.305 | 16.074 | 20.550 | 21.438 | 19.138 | 11.006 | 10.720 |
| **(010)** | 3.768 | 4.381 | 5.400 | 7.330 | 11.058 | 12.620 | 12.933 | 12.324 | 6.141 | 3.328 |
| **(001)** | 3.021 | 3.720 | 4.544 | 5.455 | 6.363 | 7.638 | 7.673 | 7.160 | 5.421 | 1.050 |

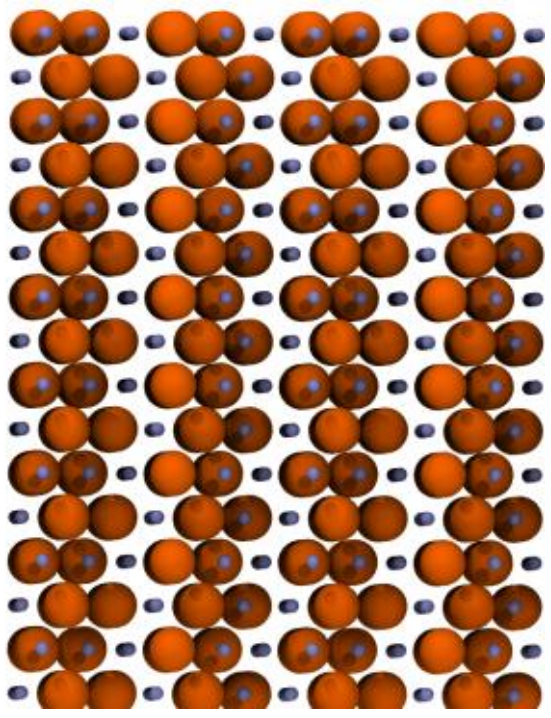

**Figure S4.** The ribbonlike morphology of $Sb_2S_3$ crystal structure.

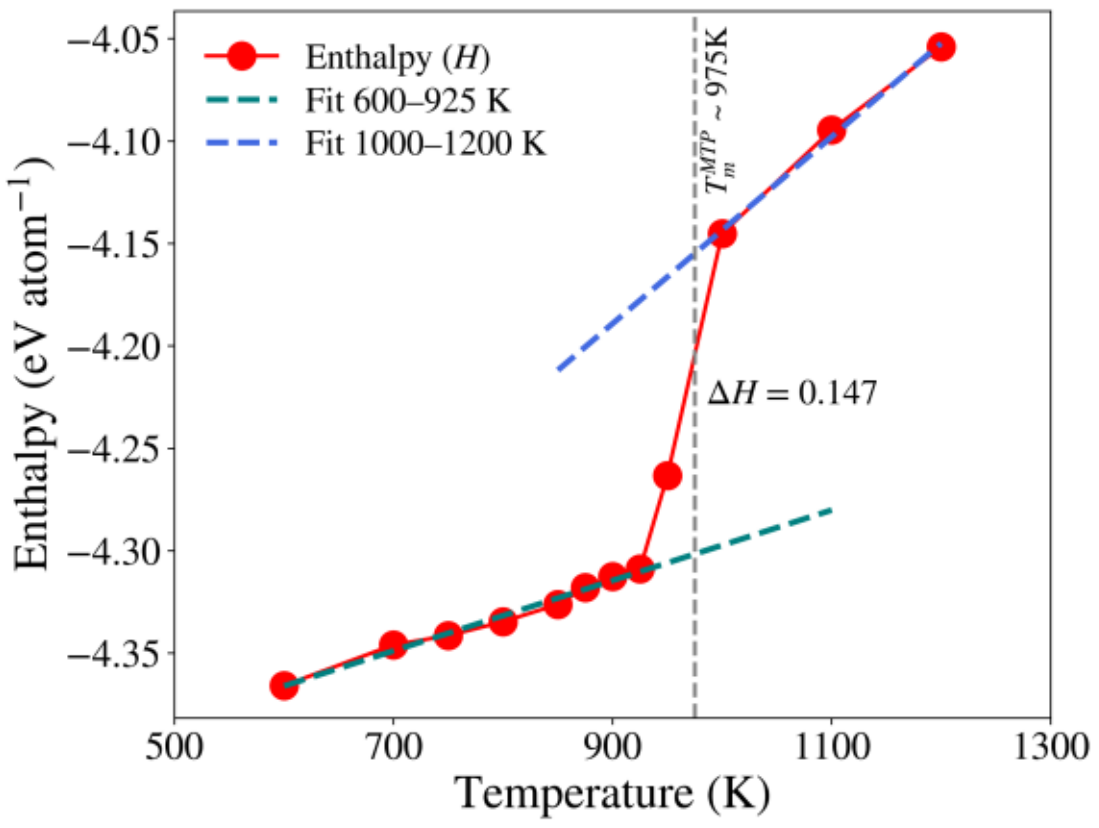

**Figure S5.** Estimation of melting point ($T_m^{MTP}$) by calculating enthalpy (facet-averaged) at different temperatures under study with the dual phase coexistence model (7680 atoms). Enthalpy was computed from last 2 ns of NPT simulation trajectory. The melting point was found to be ~975 K and enthalpy of melting (Δ*H*) was estimated to be 0.147 eV/atom.

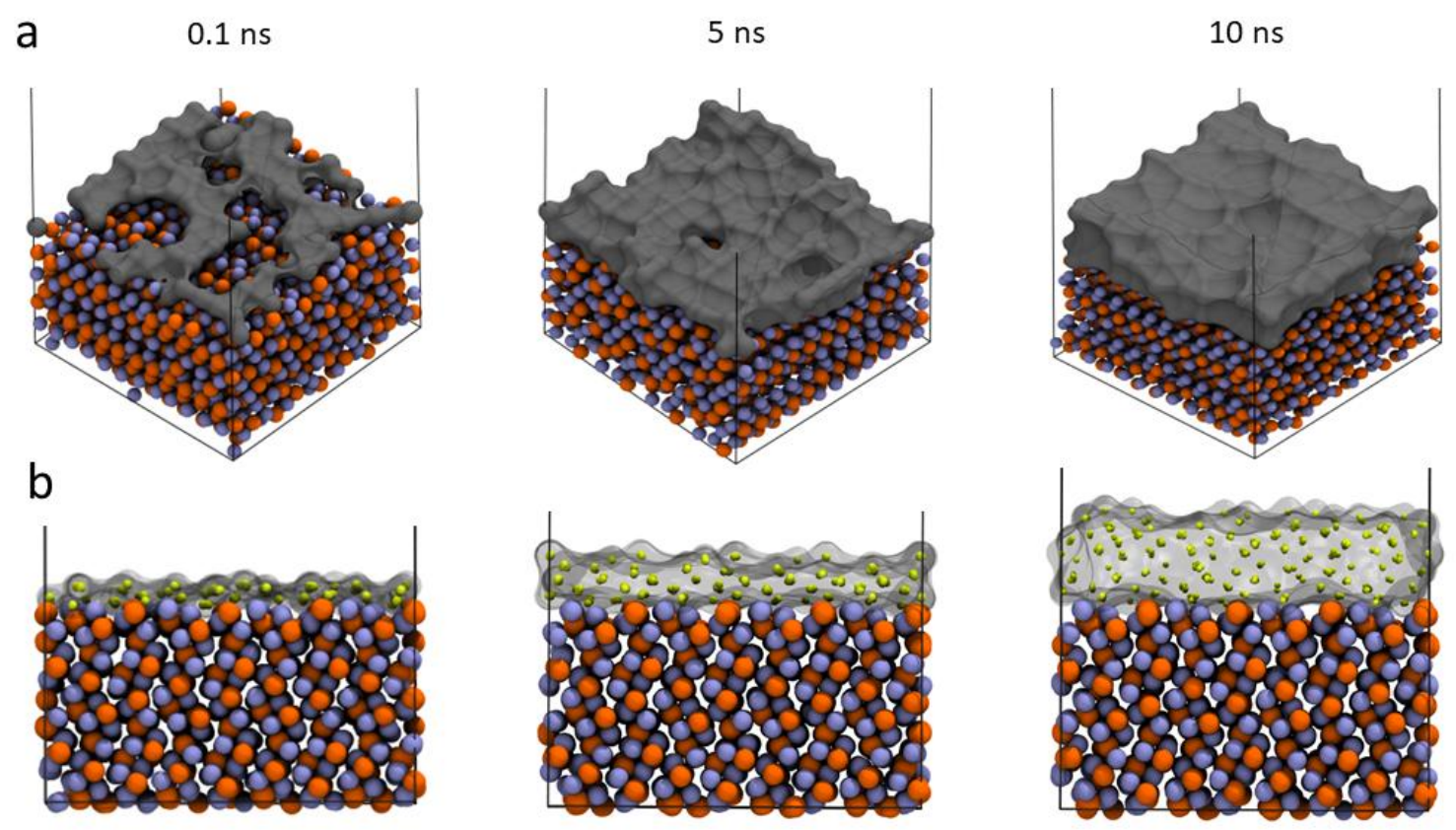

**Figure S6.** Visual representation of evolution of crystal growth front at different simulation time a) top view b) side view. The other interface is not shown for clarity. The growing layer is coloured in grey to distinguish it from the crystal slab. In (b) the crystal-like atoms are represented as yellow spheres.

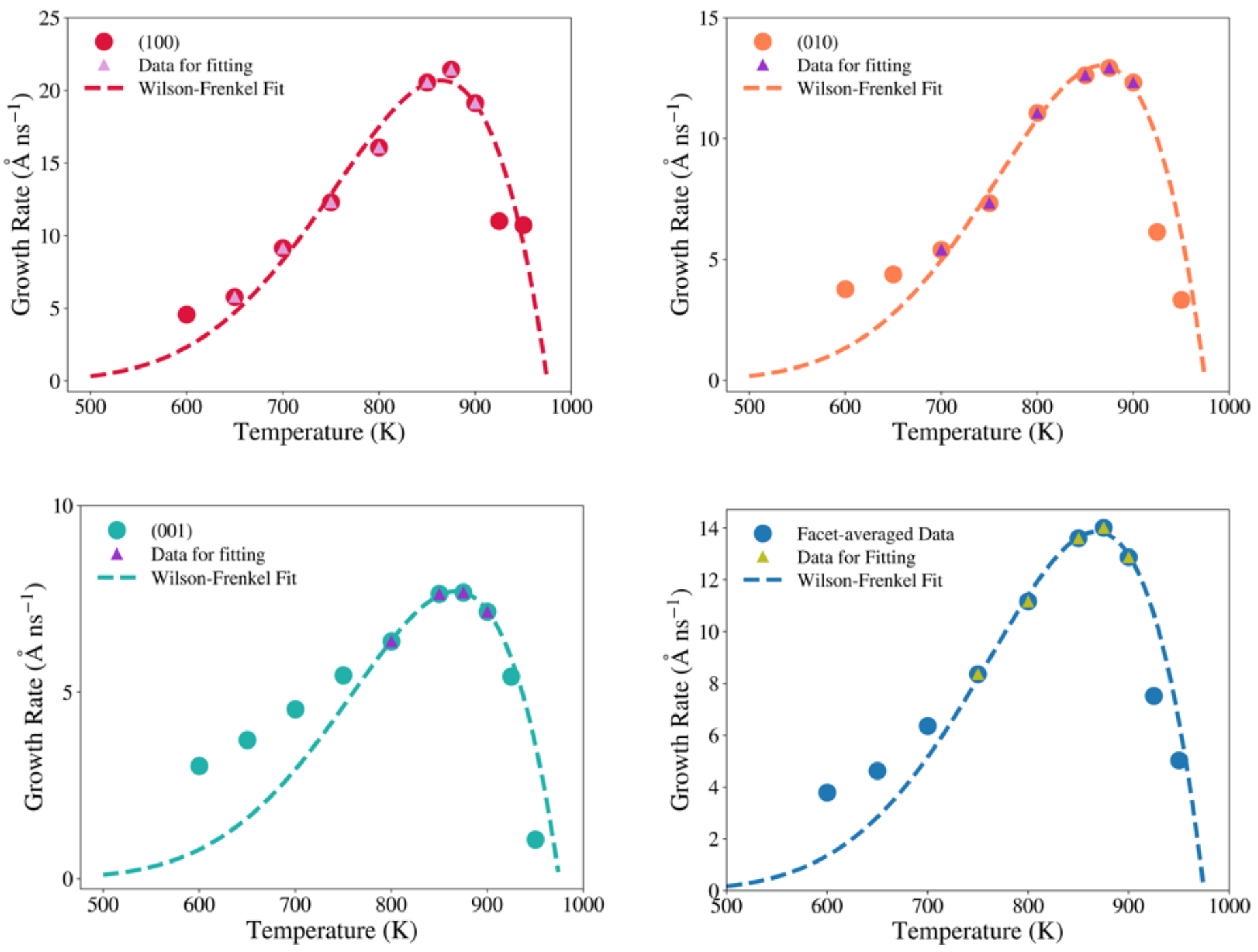

**Figure S7.** Wilson-Frenkel fitting for (100), (010), (001), and facet averaged growth rate data. The data points for fitting were selected by evaluating the best $R^2$ (coefficient of determination) values. The fitting parameters are presented in Table 3 (SI).

**Table S2:** Wilson-Frenkel fitting parameters

| **Facet** | $\Delta E_a^{WF}$ (eV) | $k_0$ | $R^2$ |
|---|---|---|---|
| **(100)** | 0.554±0.022 | 184820.919 | 0.977 |
| **(010)** | 0.570±0.015 | 144798.492 | 0.988 |
| **(001)** | 0.571±0.004 | 87188.711 | 0.997 |
| **Facet-averaged** | 0.578±0.010 | 170754.918 | 0.993 |

**Table S3:** Activation energy for crystallization for some important PCMs

| **Material** | **Crystallization Activation Energy ($E_c$)** | **Method** | **Citation (SI)** |
|---|---|---|---|
| $Sb_2S_3$ | 2 eV | Reflectivity, Kissinger Fit | Ref. 5 |
| | 1.31-2.72 eV | DSC, Kissinger Fit | Ref. 6 |
| | 0.58 eV (growth) | MD (MTP), Wilson-Frenkel Fit, Facet avg. | This study |
| $Ge_2Sb_2Te_5$ | 0.267 eV | Self-Diffusion (MD/NNP) | Ref. 4 |
| | 2.34 eV | *ab-initio* | Ref. 7 |
| AIST | 3.06 eV (low heating rate)<br>0.89 eV (high heating rate) | DSC, Kissinger Fit | Ref. 8 |
| $Sb_2Se_3$ | 0.39 eV | Temperature-dependent admittance spectroscopy (TAS) | Ref. 9 |
| GeTe | 1.7-3.9 eV | DSC, TEM, XPS, Reflectivity | Ref. 10-14 |

**Table S4:** Self-diffusion constant (D) values calculated from mean square displacement (MSD) for all systems under study

| Self-diffusion constant (D) (Å$^2$ ps$^{-1}$) at different temperatures (K) | | | | | | | | | | |
|---|---|---|---|---|---|---|---|---|---|---|
| **Facet** | **600** | **650** | **700** | **750** | **800** | **850** | **875** | **900** | **925** | **950** |
| **(100)** | 5.583e-04 | 1.323e-03 | 1.340e-03 | 1.671e-03 | 1.914e-03 | 3.529e-03 | 6.231e-03 | 8.665e-03 | 3.320e-02 | 8.039e-02 |
| **(010)** | 6.882e-04 | 1.424e-03 | 1.317e-03 | 2.759e-03 | 4.804e-03 | 4.682e-03 | 2.092e-02 | 2.305e-02 | 3.477e-02 | 6.913e-02 |
| **(001)** | 4.855e-04 | 8.500e-04 | 1.454e-03 | 3.577e-03 | 6.829e-03 | 10.949e-03 | 2.852e-02 | 5.448e-02 | 5.700e-02 | 7.739e-02 |

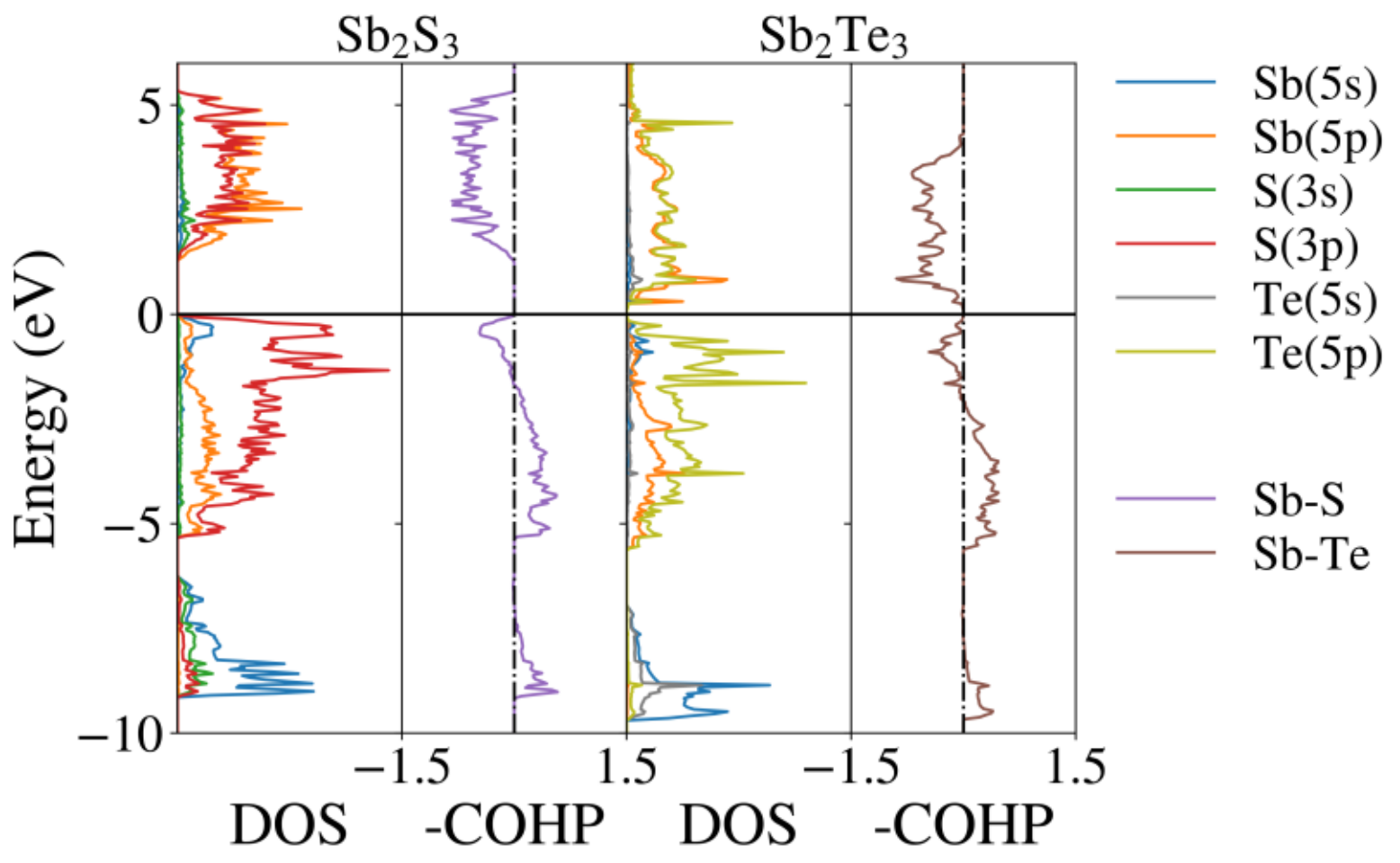

**Figure S8.** Crystal Orbital Hamilton Population (COHP) analysis for crystalline $Sb_2S_3$ and $Sb_2Te_3$.